\documentclass[12pt]{article}%

\usepackage{amssymb}
\usepackage{amsfonts}
\usepackage{amsmath}%
\usepackage{graphicx}
\usepackage{caption}
\usepackage{here}

\usepackage{amsthm}
\usepackage{ amscd}

\usepackage{euscript}
\usepackage[all,tips]{xy}

\usepackage{color}

\providecommand{\U}[1]{\protect\rule{.1in}{.1in}}

\begin{document}

\title{The $DKP$ equation in the Woods-Saxon potential well: Bound states}
\author{Boutheina  Boutabia-Ch\'{e}raitia\\Laboratoire de Probabilit\'{e}s et Statistiques (LaPS)\\Universit\'{e} Badji-Mokhtar. Annaba. Alg\'{e}rie\\\thanks{Permanent adress.} Facult\'{e} de M\'{e}decine d'Annaba. \\bboutheina@hotmail.com
\and Abdenacer  Makhlouf\\Laboratoire de Math\'{e}matiques, Informatique,\\et Applications (LMIA)\\Universit\'{e} de Haute Alsace\\Mulhouse. France\\Abdenacer.Makhlouf@uha.fr}
\maketitle

\begin{abstract}
We solve the Duffin-Kemmer-P\'{e}tiau equation in the presence of a spatially
one-dimensional symmetric potential well. We compute the scattering state
solutions and we derive conditions for transmission resonances. The bound
solutions are derived by a graphic study and the appearance of the
antiparticle bound state is discussed.



\end{abstract}

\section{Introduction}

The presence of strong fields introduces quantum phenomena, such as
supercriticallity and spontaneous pair production which is one of the most
interesting non-perturbative phenomena associated with the charged quantum vacuum.

During the last decades, a great effort has been made in understanding quantum
processes in strong fields. The discussion of overcritical behavior of bosons
requires a full understanding of the single particle spectrum, and
consequently of the exact solutions to the Klein-Gordon $(KG)$ equation.
Recently, transmission resonances for the $KG$ $\left[  1\right]  $\ and
Duffin-Kemmer-P\'{e}tiau $\left(  DKP\right)  $ equation $\left[
2,3,4\right]  $ in the presence of a Woods-Saxon $\left(  WS\right)  $
potential barrier have been computed. The transmission coefficient as a
function of the energy and the potential amplitude shows a behavior that
resembles the one obtained for the Dirac equation in $[5]$.

The $KG$ equation in the $WS$ potential well $\left[  6\right]  $ was solved
and it was shown that analogous to the square well potential, there is a
critical point $V_{cr}$ where the bound antiparticle mode appears to coalesce
with the bound particle.

In the present article, we solve the $DKP$ equation in the $WS$ potential well
and we make a graphical study for the resonance transmissions. Among the
advantages of working with the $WS$ potential we have to mention that, in the
one-dimensional case, the $DKP$ equation as well as the $KG$ and Dirac
equations are solvable in terms of special functions and therefore the study
of bound states and scattering processes becomes more tractable. We show that
the antiparticle bound states arise for the $WS$ potential well, which is a
smoothed out form of the square well. The interest in computing bound states
and spontaneous pair creation processes in such potentials lies in the fact
that they possess properties that could permit us to determine how the shape
of the potential affects the pair creation mechanism.

The article is structured as follows: Section $2$ is devoted to solving the
$DKP$ equation in the presence of the one dimensional $WS$ potential well. In
Section $3$ We derive the equation governing the eigenvalues corresponding to
the bound states and compute the bound states. Finally, in Section$4$, we
briefly summarize our results.

\section{The $DKP$ equation in the $WS$ potential well}

The $DKP$ equation $\left[  7,8,9\right]  $ is a natural manner to extend the
covariant Dirac formalism to the case of scalar (spin $0$) and vectorial (spin
$1$) particles when interacting with an electromagnetic field. It will be
written as $\left(  \hbar=c=1\right)  $:%

\begin{equation}
\left[  i\beta^{\mu}\left(  \partial_{\mu}+ieA_{\mu}\right)  -m\right]
\psi\left(  \mathbf{r},t\right)  =0
\end{equation}
where the matrices $\beta^{\mu}$ verify the $DKP$ algebra:%

\begin{equation}
\beta^{\mu}\beta^{\nu}\beta^{\lambda}+\beta^{\lambda}\beta^{\nu}\beta^{\mu
}=g^{\mu\nu}\beta^{\lambda}+g^{\nu\lambda}\beta^{\mu}%
\end{equation}
where the convention for the metric tensor is here\textbf{\ }$g^{\mu\nu
}=diag\left(  1,-1,-1,-1\right)  .$ The algebra $\left(  2\right)  $ has three
irreductible representations whose degrees are $1,5$, and $10$. The first one
is trivial, having no physical content, the second and the third ones
correspond respectively to the scalar and vectorial representations. For the
spin $0$, the $\beta^{\mu}$\ are given by$:$%
\begin{equation}%
\begin{array}
[c]{ll}%
\beta^{0}=\left(
\begin{array}
[c]{cc}%
\mathbf{\theta} & \mathbf{0}\\
\mathbf{0} & \mathbf{0}%
\end{array}
\right)  ; & \beta^{i}=\left(
\begin{array}
[c]{cc}%
\mathbf{0} & \rho^{i}\\
-\rho_{T}^{i} & \mathbf{0}%
\end{array}
\right)  ;i=1,2,3
\end{array}
\end{equation}
\ with
\begin{equation}%
\begin{array}
[c]{ll}%
\rho^{1}=\left(
\begin{array}
[c]{ccc}%
-1 & 0 & 0\\
0 & 0 & 0
\end{array}
\right)  , & \rho^{2}=\left(
\begin{array}
[c]{ccc}%
0 & -1 & 0\\
0 & 0 & 0
\end{array}
\right) \\
\rho^{3}=\left(
\begin{array}
[c]{ccc}%
0 & 0 & -1\\
0 & 0 & 0
\end{array}
\right)  , & \mathbf{\theta}=\left(
\begin{array}
[c]{cc}%
0 & 1\\
1 & 0
\end{array}
\right)
\end{array}
\end{equation}
the $\rho_{T}$\ denoting the transposed matrix of $\rho,$\ and $\mathbf{0}%
$\ the zero matrix. For the spin $1,$\ the $\beta^{\mu}$\ are given by:%
\begin{equation}%
\begin{array}
[c]{ll}%
\beta^{0}=\left(
\begin{array}
[c]{cccc}%
0 & \overline{0} & \overline{0} & \overline{0}\\
\overline{0}^{T} & \mathbf{0} & \mathbf{1} & \mathbf{0}\\
\overline{0}^{T} & \mathbf{1} & \mathbf{0} & \mathbf{0}\\
\overline{0}^{T} & \mathbf{0} & \mathbf{0} & \mathbf{0}%
\end{array}
\right)  ; & \beta^{i}=\left(
\begin{array}
[c]{cccc}%
0 & \overline{0} & e_{i} & \overline{0}\\
\overline{0}^{T} & \mathbf{0} & \mathbf{0} & -is_{i}\\
-e_{i}^{T} & \mathbf{0} & \mathbf{0} & \mathbf{0}\\
\overline{0}^{T} & -is_{i} & \mathbf{0} & \mathbf{0}%
\end{array}
\right)  ;\text{\ }i=1,2,3
\end{array}
\end{equation}
with%
\begin{equation}
e_{1}=\left(  1,0,0\right)  ;e_{2}=\left(  0,1,0\right)  ;e_{3}=\left(
0,0,1\right)  ;\overline{0}=\left(  0,0,0\right)
\end{equation}
$\mathbf{0}$\ and $\mathbf{1}$\ denoting respectively the zero matrix and the
unity matrix, and the $s_{i}$\ being the standard nonrelativistic $(3\times
3)$\ spin $1$\ matrices:%
\begin{equation}
s_{1}=\left(
\begin{array}
[c]{lll}%
0 & 0 & 0\\
0 & 0 & -i\\
0 & i & 0
\end{array}
\right)  ,s_{2}=\left(
\begin{array}
[c]{lll}%
0 & 0 & i\\
0 & 0 & 0\\
-i & 0 & 0
\end{array}
\right)  ,s_{3}=\left(
\begin{array}
[c]{lll}%
0 & -i & 0\\
i & 0 & 0\\
0 & 0 & 0
\end{array}
\right)
\end{equation}

The $DKP$\ particles we consider are in interaction with the $WS$ potential
defined by:
\begin{equation}
V_{r}\left(  z\right)  =\dfrac{-V_{0}}{1+\exp\left(  \frac{\left\vert
z\right\vert -a}{r}\right)  }%
\end{equation}
where $V_{0}$ is real and positive, $a>0$ and $r>0$ are real, positive and
adjustable.\textbf{\ }

The form of the $WS$ potential is shown in the Fig$.1$, from which one readily
notices that for a given value of the width parameter $a$, as the shape
parameter $r$ decreases $\left(  r\longrightarrow0^{+}\right)  $, the $WS$
potential reduces to a square well with smooth walls:%

\begin{figure}[h]
\captionsetup{justification=centering}
  \centering\includegraphics[scale=0.6]{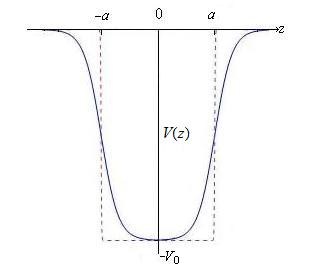}
   \centering\caption{
\small{The $WS$ potential for\\  $a=2$, with $r=\frac{1}{3}$ (solide line)\\ 
and 
$r=\frac{1}{100}$ (dotted line) }}
\end{figure}

\[
V\left(  z\right)  =-V_{0}\theta\left(  a-\left\vert z\right\vert \right)
\]%
\begin{equation}
=\left\{
\begin{array}
[c]{c}%
-V_{0}\text{ \ \ for }\left\vert z\right\vert \leqslant a\\
0\text{\ \ \ for }\left\vert z\right\vert >a
\end{array}
\right.
\end{equation}

The interaction being scalar and independent of time, one can choose for
$\psi(z,t)$\ the form $e^{-iEt}\kappa\left(  z\right)  ,$\ so one gets the
following eigenvalue equation:
\begin{equation}
\left[  \beta^{0}\left(  E-eV\right)  +i\beta^{3}\frac{d}{dz}-m\right]
\kappa\left(  z\right)  =0
\end{equation}
with $\kappa\left(  z\right)  ^{T}=\left(  \varphi,\mathbf{A},\mathbf{B}%
,\mathbf{C}\right)  ,$ $\mathbf{A,B}$ and $\mathbf{C}$ being respectively
vectors of components $A_{i},B_{i}$ and $C_{i};$ $i=1,2,3.$ According to the
equations they satisfy, one gathers the components of $\kappa\left(  z\right)
$ this way%
\begin{equation}
\Psi^{T}=\left(  A_{1},A_{2},B_{3}\right)  ,\Phi^{T}=\left(  B_{1},B_{2}%
,A_{3}\right)  ,\Theta^{T}=\left(  C_{2},-C_{1},\varphi\right)  ;\text{ and
}C_{3}=0
\end{equation}
with
\begin{equation}
\mathbf{O}_{KG}\Psi=0
\end{equation}%
\begin{equation}
\left(
\begin{array}
[c]{c}%
\Phi\\
\Theta
\end{array}
\right)  =\left(
\begin{array}
[c]{c}%
\frac{E-eV}{m}\\
\frac{i}{m}\frac{d}{dz}%
\end{array}
\right)  \otimes\Psi
\end{equation}
then one will designate by $\phi\left(  z\right)  ^{T}=\left(  \Psi
,\Phi,\Theta\right)  $\ the solution of $\left(  10\right)  \left[  2\right]
,$\ \newline$O_{KG}=\frac{d^{2}}{dz^{2}}+\left[  \left(  E-eV\right)
^{2}-m^{2}\right]  $\ being the Klein-Gordon $"KG"$\ operator.\textbf{\ }

By the following, one will follow the same steps as for the barrier potential
$\left[  2\right]  $, where one will replace $V_{0}$\ by $-V_{0}$, then one
gets the asymptotic behavior of the wave function at $\left\vert z\right\vert
\longrightarrow\infty:$%
\begin{equation}
\left(
\begin{array}
[c]{c}%
\Psi\\
\Omega\\
\Theta
\end{array}
\right)  \underset{z\longrightarrow-\infty}{\longrightarrow}Ae^{-ik\left(
z+a\right)  }\left(
\begin{array}
[c]{c}%
1\\
\dfrac{E}{m}\\
-\dfrac{i\mu}{rm}%
\end{array}
\right)  \otimes V+Be^{ik\left(  z+a\right)  }\left(
\begin{array}
[c]{c}%
1\\
\dfrac{E}{m}\\
\dfrac{i\mu}{rm}%
\end{array}
\right)  \otimes V
\end{equation}

\begin{equation}
\left(
\begin{array}
[c]{c}%
\Psi\\
\Omega\\
\Theta
\end{array}
\right)  \underset{z\longrightarrow+\infty}{\longrightarrow}Ce^{ik\left(
z-a\right)  }\left(
\begin{array}
[c]{c}%
1\\
\dfrac{E}{m}\\
\dfrac{i\mu}{rm}%
\end{array}
\right)  \otimes V+De^{-ik\left(  z-a\right)  }\left(
\begin{array}
[c]{c}%
1\\
\dfrac{E}{m}\\
-\dfrac{i\mu}{rm}%
\end{array}
\right)  \otimes V
\end{equation}
with the following definitions of the coefficients:

$B$ and $D$ are respectively the coefficients of the incoming waves from
$-\infty\longrightarrow0$ and from $+\infty\longrightarrow0.$

$A$ and $C$ are respectively the coefficients of the reflected and transmitted wave.

$V$ is a constant vector of dimension $\left(  3\times1\right)  :$
\begin{equation}
\mathbf{V}=\left(
\begin{array}
[c]{c}%
N_{1}\\
N_{2}\\
N_{3}%
\end{array}
\right)
\end{equation}

The coefficients of reflection $\mathbf{R}$ and transmission $\mathbf{T}$ will
be given by $\left[  2\right]  :$%
\begin{equation}
\mathbf{R}=\frac{1}{4}\left\vert \lambda^{2\mu}\right\vert ^{2}\left\vert
\frac{F_{6}}{F_{5}}+\frac{F_{2}}{F_{1}}\right\vert ^{2}%
\end{equation}
and
\begin{equation}
\mathbf{T}=\frac{1}{4}\left\vert \lambda^{2\mu}\right\vert ^{2}\left\vert
\frac{F_{6}}{F_{5}}-\frac{F_{2}}{F_{1}}\right\vert ^{2}%
\end{equation}
with%
\begin{equation}%
\begin{tabular}
[c]{l}%
$\lambda=\frac{1}{1+\exp(-\frac{a}{r})}$\\
$F_{1}=\text{ }_{2}F_{1}\left(  \alpha_{1},\beta_{1},\gamma_{1},\lambda
\right)  $\\
$F_{2}=\text{ }_{2}F_{1}\left(  \alpha_{2},\beta_{2},\gamma_{2},\lambda
\right)  $\\
$F_{3}=\text{ }_{2}F_{1}\left(  \alpha_{1}+1,\beta_{1}+1,\gamma_{1}%
+1,\lambda\right)  $\\
$F_{4}=\text{ }_{2}F_{1}\left(  \alpha_{2}+1,\beta_{2}+1,\gamma_{2}%
+1,\lambda\right)  $\\
$F_{5}=\left[  -\mu+\lambda\left(  \mu-\nu\right)  \right]  F_{1}%
+\lambda\left(  1-\lambda\right)  \frac{\alpha_{1}\beta_{1}}{\gamma_{1}}F_{3}%
$\\
$F_{6}=\left[  \mu-\lambda\left(  \mu+\nu\right)  \right]  F_{2}%
+\lambda\left(  1-\lambda\right)  \frac{\alpha_{2}\beta_{2}}{\gamma_{2}}F_{4}$%
\end{tabular}
\end{equation}
and%
\begin{equation}
\left\{
\begin{array}
[c]{l}%
\alpha=\left(  \mu+\nu+\frac{1}{2}\right)  -\dfrac{\nu_{0}}{2}\\
\beta=\left(  \mu+\nu+\frac{1}{2}\right)  +\dfrac{\nu_{0}}{2}\\
\gamma=1+2\mu\\
\mu^{2}=r^{2}\left(  m^{2}-E^{2}\right)  ,\mu=irk\text{ }\\
\nu^{2}=r^{2}\left[  m^{2}-\left(  E+eV_{0}\right)  ^{2}\right]  ,\text{ }%
\nu=irp\text{ \ with }p\text{ real }\\
\nu_{0}=\sqrt{\left(  1-2reV_{0}\right)  \left(  1+2reV_{0}\right)  }%
\end{array}
\right.
\end{equation}

Remark that one will distinguish two cases: $\left\vert E\right\vert >m,$ i.e.
$k$ is real, which solutions are called scattering states $\left[
2,3,4\right]  $, and $\left\vert E\right\vert <m,$ i.e. $k$ is imaginary and
which solutions are bound states.
\begin{equation}
\left\{
\begin{array}
[c]{l}%
\alpha_{1}=\left(  -\mu+\nu+\frac{1}{2}\right)  -\frac{\nu_{0}}{2}\\
\beta_{1}=\left(  -\mu+\nu+\frac{1}{2}\right)  +\frac{\nu_{0}}{2}\\
\gamma_{1}=1-2\mu
\end{array}
\right.  ,\left\{
\begin{array}
[c]{l}%
\alpha_{2}=\left(  \mu+\nu+\frac{1}{2}\right)  -\frac{\nu_{0}}{2}\\
\beta_{2}=\left(  \mu+\nu+\frac{1}{2}\right)  +\frac{\nu_{0}}{2}\\
\gamma_{2}=1+2\mu
\end{array}
\right.
\end{equation}

\section{Bound states}

To get the transmission coefficient for bound states in terms of $E$ and of
$\left(  -eV_{0}\right)  ,$ we proceed to solve numerically the equation
$(18).$ When varying the depth $\left(  -eV_{0}\right)  $ of the square well,
we get:%

\begin{figure}[H]

  \begin{minipage}[b]{0.5\linewidth}
   \centering
   \includegraphics[width=6cm,height=4cm]{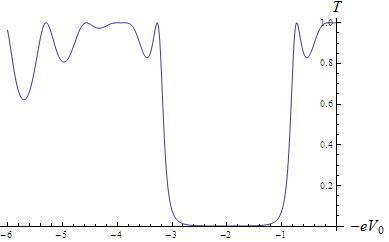}      
  \centering \caption{\small{The $\textbf{T}$ coefficient in\\  terms of the depth $\left(
-eV_0\right)  $ \\ of the potential,  for\\  $a=2,$ $E=-2m$  and $m=1$ }}
  \end{minipage}
  \begin{minipage}[b]{0.48\linewidth}
   \centering
   \includegraphics[width=6cm,height=4cm]{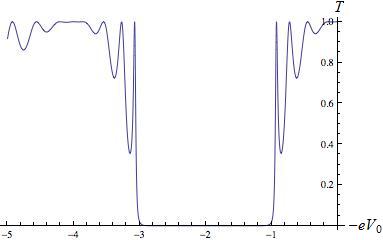}    
   \centering\caption{\small{ The $\textbf{T}$ coefficient in \\  terms of the depth $\left(
-eV_0\right)  $ \\ of the potential, for\\  $a=4,$ $E=-2m$  and $m=1$ }}  
  \end{minipage}

\end{figure}

and when varying the energy $E$ of the particle, we get:%

\begin{figure}[H]
  \begin{minipage}[b]{0.5\linewidth}
   \centering
   \includegraphics[width=6cm,height=4cm]{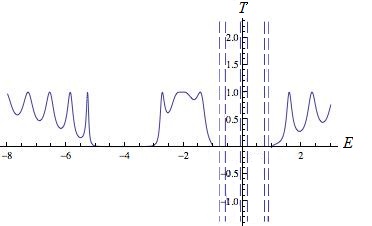}    
   \centering\caption{
\small{The $\textbf{T}$ coefficient in\\  terms of the energy $E$
for \\ $m=1,eV_0=4,a=2.$ The\\  energies of the bound states are\\  depicted by dashed
lines. }}  
  \end{minipage}
  \begin{minipage}[b]{0.48\linewidth}
   \centering
   \includegraphics[width=6cm,height=4cm]{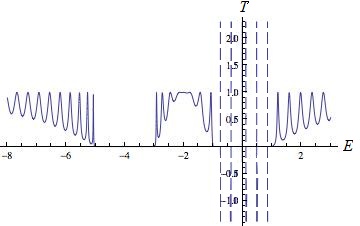}    
   \centering\caption{
\small{The $\textbf{T}$ coefficient in \\ terms of the energy $E$
for \\ $m=1,eV_0=4,a=4.$ \\ The energies of the bound states \\ are depicted by dashed
lines.} }  
  \end{minipage}
\end{figure}

Fig. $2$ and Fig. $3$ show that as in the Dirac $\left[  10\right]
$\textbf{\ }and the $KG$ cases $\left[  6\right]  $, the transmission
coefficient vanishes for values of the potential strength $E-m<-eV_{0}<E+m,$
and transmission resonances appear for $V_{0}>E+m$. They also show that the
width of the transmission resonances decreases as the parameter $a$ decreases.

Fig. $4$ and Fig. $5$, show that the occurrence of the transmission resonances
increases with the width $a$ of the square well. As in the case of the Dirac
particle $\left[  10\right]  ,$ we have significant structures of resonance in
the particle continuum for $E>m,$ and in the antiparticle continuum for
$E<-eV_{0}-m$. Antiparticles with lower\textbf{\ }energy $(-eV_{0}%
-m<E<-eV_{0}+m)$ only penetrate the well with a probability which decreases
with the width $a$ of the well. Hence\textbf{\ }$\mathbf{T}$\textbf{\ }is
about zero in our case.\textbf{\ }However, in the domain $-eV_{0}%
+m<E<-m\left(  \Longrightarrow eV_{0}>2m\right)  $, there is the possibility
that the incoming wave meets a bound state, and thus penetrates the potential
domain more or less unhindered. At the point where by extrapolation of the
spectrum of the bound states (see Fig. $6$), one would expect the quasi
bound state, $\mathbf{T}$ is equal to $1$. The dived bound state in this way
becomes perceptible as a resonance in the scattering spectrum below $E=-m$.

\begin{figure}[h]
\captionsetup{justification=centering}
  \centering\includegraphics[scale=1.]{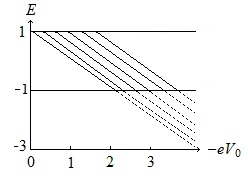}
   \centering\caption{
\small{Eigenvalue spectrum \\ for $m=1,a=4.$} }
\end{figure}

In Fig. $6$, the energies of the dived states corresponding to resonances
are depicted by dashed lines. They are extracted from the maxima of the
transmission coefficients of Fig. $5$

By the following, one wants to get the dependence of the spectrum of bound
states (i.e. $\left\vert E\right\vert <m)$ on the potential strength $V_{0}.$
One uses for this aim, the unitary condition that coefficients $\mathbf{R}$
and $\mathbf{T}$ verify, i.e. $\mathbf{R+T=}1,\ $\newline which leads to:%
\begin{equation}
1\mathbf{=}\frac{1}{2}\left\vert \lambda^{2\mu}\right\vert ^{2}\left[
\left\vert \frac{F_{6}}{F_{5}}\right\vert ^{2}+\left\vert \frac{F_{2}}{F_{1}%
}\right\vert ^{2}\right]
\end{equation}
So%
\[
\frac{2}{\left\vert \lambda^{2\mu}\right\vert ^{2}}=
\]%
\[
\left\vert \frac{\left[  \mu-\lambda\left(  \mu+\nu\right)  \right]  _{2}%
F_{1}\left(  \alpha_{2},\beta_{2},\gamma_{2},\lambda\right)  +\lambda\left(
1-\lambda\right)  \frac{\alpha_{2}\beta_{2}}{\gamma_{2}}_{2}F_{1}\left(
\alpha_{2}+1,\beta_{2}+1,\gamma_{2}+1,\lambda\right)  }{\left[  -\mu
+\lambda\left(  \mu-\nu\right)  \right]  _{2}F_{1}\left(  \alpha_{1},\beta
_{1},\gamma_{1},\lambda\right)  +\lambda\left(  1-\lambda\right)  \frac
{\alpha_{1}\beta_{1}}{\gamma_{1}}_{2}F_{1}\left(  \alpha_{1}+1,\beta
_{1}+1,\gamma_{1}+1,\lambda\right)  }\right\vert ^{2}%
\]%
\begin{equation}
+\left\vert \frac{_{2}F_{1}\left(  \alpha_{2},\beta_{2},\gamma_{2}%
,\lambda\right)  }{_{2}F_{1}\left(  \alpha_{1},\beta_{1},\gamma_{1}%
,\lambda\right)  }\right\vert ^{2}%
\end{equation}
One proceeds to solve\ numerically the equation $\left(  23\right)  $\ and
thus one determines the energy spectrum of the bound solutions for several
sets of parameters $r$ and $a,$ using the Gauss hypergeometric function:%
\[
F\left(  \alpha,\beta,\gamma,z\right)  =\frac{\Gamma\left(  \gamma\right)
\Gamma\left(  \beta-\alpha\right)  }{\Gamma\left(  \beta\right)  \Gamma\left(
\gamma-\alpha\right)  }\left(  -z\right)  ^{-\alpha}\times\frac{\Gamma\left(
\alpha-\beta+1\right)  \Gamma\left(  \gamma-\alpha-\beta\right)  }%
{\Gamma\left(  1-\beta\right)  \Gamma\left(  \gamma-\beta\right)  }%
\]
\
\begin{equation}
+\frac{\Gamma\left(  \gamma\right)  \Gamma\left(  \alpha-\beta\right)
}{\Gamma\left(  \alpha\right)  \Gamma\left(  \gamma-\beta\right)  }\left(
-z\right)  ^{-\beta}\times\frac{\Gamma\left(  \beta-\alpha+1\right)
\Gamma\left(  \gamma-\beta-\alpha\right)  }{\Gamma\left(  1-\alpha\right)
\Gamma\left(  \gamma-\alpha\right)  }\text{ \ \ \ }\left\vert \arg
(-z)\right\vert <\pi\text{\ }%
\end{equation}
For that, one puts $\lambda_{\epsilon}=\lambda-i\varepsilon,$ then:
\[
F\left(  \alpha,\beta,\gamma,\lambda\right)  =\underset{\epsilon\rightarrow
0}{\lim}F\left(  \alpha,\beta,\gamma,\lambda_{\epsilon}\right)
\]
so one obtains:

\begin{figure}[H]

  \begin{minipage}[b]{0.5\linewidth}
   \centering
   \includegraphics[width=6cm,height=6cm]{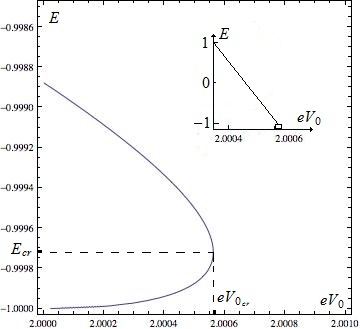}    
   \centering\caption{
\small{Bound state spectrum \\ for $m=1,a=1,r=0.00015$} }  
  \end{minipage}
  \begin{minipage}[b]{0.48\linewidth}
   \centering
   \includegraphics[width=6cm,height=6cm]{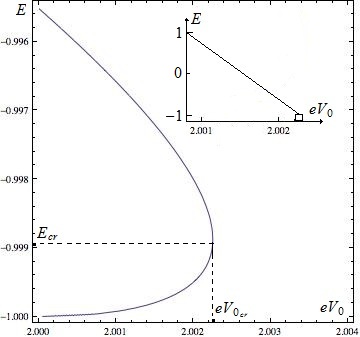}     
   \centering\caption{
\small{Bound state spectrum \\ for $m=1,a=4,r=0.0003$} } 
  \end{minipage}
\end{figure}

In Fig. $7\left(  \text{resp. Fig. }8\right)  $, the bound state for the
antiparticle appears for the domain $2.0004<eV_{0}<2.0006\left(  \text{resp.
}2.002<eV_{0}<2.003\text{ }\right)  $. For $eV_{0}\simeq2.00055m\left(
\text{resp. }eV_{0}\simeq2.0022m\right)  $, the well becomes supercritical.
The lowest bound state enters the lower continuum and can there be realized as
a resonance in the transmission coefficient. This critical value is depicted
by $\left(  eV_{0}\right)  _{cr},$ and its corresponding energy by $E_{cr}.$
The appearance of these bound antiparticle states is corresponding with the
short range of the potential.

\begin{figure}[H]

  \begin{minipage}[b]{0.5\linewidth}
   \centering
   \includegraphics[width=6cm,height=4cm]{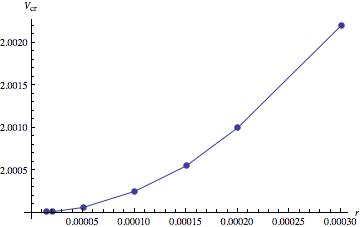}    
   \centering\caption{
{\small{ Critical potential ${eV_0}_cr$ \\ versus the shape parameter $r$ for\\ 
$m=1,a=4$ } }}  
  \end{minipage}
  \begin{minipage}[b]{0.48\linewidth}
   \centering
   \includegraphics[width=6cm,height=4cm]{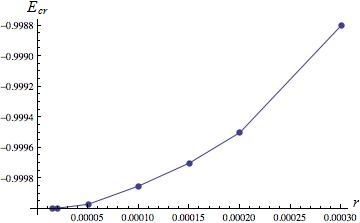}   
      \centering\caption{
\small{Critical energy $E_cr$ \\ versus the shape parameter $r$ for\\ 
$m=1,a=4$} }   
  \end{minipage}

  \end{figure}

Fig. $9$ shows that when the shape parameter $r$ increases, the critical
potential value $eV_{0_{cr}}$ where the bound antiparticle mode appears to
coalesce with the bound particle increases.

Fig. $10$ shows that as in the $KG$ case $\left[  6\right]  ,$ when the shape
parameter $r$ increases, the critical energy value $E_{cr}$ for which
antiparticle state appears increases.

\begin{figure}[!h]
\captionsetup{justification=centering}
  \centering\includegraphics[scale=0.6]{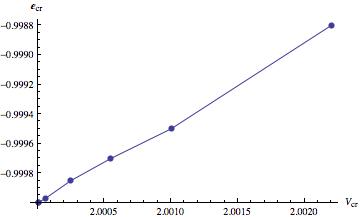}
   \centering\caption{
\small{Critical energy $E_{cr}$ versus critical potential ${eV_0}_{cr}$ \\ for $m=1,a=4$} }
\end{figure}

For $a=4$, we have moved the shape parameter $r$ from $0.000015$ to $0.0003$.
Fig. $11$ shows the behavior of the critical energy value $E_{cr}$ versus the
critical potential $eV_{0_{cr}}.$ We notice that as the value of $eV_{0_{cr}}$
increases, the energy value for which the antiparticle state appears increases.

\section{Conclusion}

We have showed a similarity in behavior between $DKP$\textbf{, }%
$KG$\textbf{\ }and Dirac particles, when interacting with a one-dimensional
potential well. The resonances being interpreted as the signature for
spontaneous pair creation, we have demonstrated that the $WS$ potential well
is able to bind particles. These resonances do not exist for subcritical potentials.

Transmission resonances for the one dimensional $DKP$ equation possesses the
same rich structure that we observe for the Dirac and the $KG$ equations. For
the $DKP$\ and $KG$\ particles, this can be interpreted as a demonstration of
the equivalence between $DKP$\ and $KG$\ theories. For $DKP$ and Dirac
particles in a one-dimensional potential well, the bound state always exists,
independent of the depth and the width of the potential. This being opposite
with the corresponding three-dimensional problem where not every potential
well has a bound state.

\end{document}